\documentclass[a4paper,11pt]{article}
\pdfoutput=1 
\usepackage{amssymb,amsmath}

\usepackage{jcappub} 
\usepackage[T1]{fontenc} 
\usepackage{newtxtext,newtxmath}
\usepackage{ae,aecompl}
\usepackage{hyperref}
\usepackage{cleveref}
\usepackage{changes}
\usepackage{slashbox}
\usepackage{enumitem}

\def\prd{Physical Review D}
\def\jcap{JCAP}

\def\apj{ApJ}
\def\apjl{ApJL}

\def\aap{A\&A}

\def\mnras{MNRAS}
\def\nat{Nature}

\def\Mch{{\cal M}}

\def\beq#1{\begin{equation}\label{#1}}
\def\eeq{\end{equation}}
\def\beqa#1{\begin{eqnarray}\label{#1}}
\def\eeqa{\end{eqnarray}}

\def\comment#1{\relax}
\def\dfrac#1#2{\displaystyle\frac{#1}{#2}}

\newcommand{\be}{\begin{eqnarray}}
\newcommand{\ee}{\end{eqnarray}}

\title{\boldmath On mass distribution of coalescing 
black holes}

\author[a,b] {A.D. Dolgov,}
\author[c] {A.G. Kuranov,}
\author[d]{N.A. Mitichkin,}
\author[a] {S. Porey,}
\author[a,b,c,1]{K.A. Postnov,\note{Corresponding author.}}
\author[c]{O.S. Sazhina,}
\author[e]{I.V. Simkin}

\affiliation[a]{Department of Physics, Novosibirsk State University, \\Pirogova 2, 630090, Novosibirsk, Russia}
\affiliation[b]{ITEP, Bol. Cheremushkinskaya 25, Moscow, 117218, Russia }
\affiliation[c]{Sternberg Astronomical Institute, M.V. Lomonosov Moscow State University,\\ 13, Universitetskij pr., 119234, Moscow, Russia}
\affiliation[d]{Faculty of Physics, M.V. Lomonosov Moscow State University,\\ Leninskie Gory, 1, 119991, Moscow, Russia}
\affiliation[e]{Bauman Moscow State Technical Unisversity, Moscow, Russia}

\emailAdd{dolgov@fe.infn.it}
\emailAdd{alexandre.kuranov@gmail.com}
\emailAdd{mitichkin.nikita99@mail.ru}
\emailAdd{shiladityamailbox@gmail.com} 
\emailAdd{pk@sai.msu.ru}
\emailAdd{cosmologia@yandex.ru}
\emailAdd{vanyasimkin@gmail.com}

\abstract{
Available data on the chirp mass distribution of the coalescing black hole binaries in O1-O3 LIGO/Virgo runs are
analyzed and compared statistically with the distribution calculated under the assumption that these black holes are primordial with a log-normal mass spectrum. The theoretically calculated chirp mass distribution with the inferred best acceptable mass spectrum parameters, $M_0=17 M_\odot$ and $\gamma=0.9$, perfectly describes the data. The value of $M_0$ very well agrees with the theoretically expected one. 
On the opposite, the chirp mass distribution of black hole binaries originated from massive binary star evolution requires additional model adjustments to reproduce the observed chirp mass distribution.
}

\begin{document}

\maketitle
\flushbottom

\section{Introduction}
\label{s:intro}

The tremendous success of gravitational-wave (GW) astronomy started after the discovery of the first coalescing binary black hole (BH) GW150914 \cite{2016PhRvL.116f1102A} 
Before the suspension of the third observing run (O3) in the end of March, 2020, the LIGO/Virgo have detected 67 GW sources \footnote{https://www.ligo.caltech.edu/LA/news/ligo20200326}. Most of the O3 detections are thought to be coalescing binary BHs \footnote{https://gracedb.ligo.org/superevents/public/O3/}. The current detection sensitivity of the  LIGO/Virgo interferometers for binary inspirals   corresponds to a detection horizon of $D_h\approx 120 [\mathrm{Mpc}]\Mch^{5/6}$ , where the chirp mass of a binary with masses of the components  $m_1$ and $m_2$ is $\Mch=(m_1m_2)^{3/5}/(m_1+m_2)^{1/5}$ (see \cite{2017arXiv170908079C} for the definition of the binary inspiral horizon).


The origin of the observed coalescing BH binaries is not fully clear. The evolution of massive binary systems \cite{2016Natur.534..512B,2016A&A...588A..50M,2018MNRAS.474.2959G} is able to reproduce the observed masses and effective spins of the LIGO BH+BH sources  \cite{2017arXiv170607053B,2017ApJ...842..111H,2018A&A...616A..28Q,2019MNRAS.483.3288P,2020arXiv200411866O}, there are alternative (or additional) mechanisms of the binary BH formation. These include, in particular, the dynamical formation of close binary BHs in dense stellar clusters \cite{2016PhRvD..93h4029R,2016ApJ...824L...8R} or coalescences of primordial black hole (PBH) binaries which can constitute a fraction of dark matter $f_\mathrm{PBH}=\Omega_\mathrm{PBH}/\Omega_{DM}$ \cite{1997ApJ...487L.139N,1998PhRvD..58f3003I,2016PhRvL.116t1301B,2016PhRvL.117f1101S,2016JCAP...11..036B,2016PhRvD..94h3504C,2016arXiv160404932E}; see \cite{2020arXiv200212778C} for a recent review. 

PBHs were first introduced by Zeldovich and Novikov \cite{1967SvA....10..602Z} and Carr and Hawking \cite{1974MNRAS.168..399C}.
Depending on the model, PBHs formed at a given epoch can have different mass spectrum, including power-law $dN/dM\sim M^{-\alpha}$ \cite{1975ApJ...201....1C}, log-normal $dN/dM\sim \exp[-\gamma\ln(M/M_0)^2]$
\cite{1993PhRvD..47.4244D}, or multipeak one \cite{2019PhRvD..99j3535C} (see  \cite{2020arXiv200602838C} for more detail and references). 

At the end of the 1990s, stellar-mass binary PBHs were proposed as possible GW sources \cite{1997ApJ...487L.139N}. The formation and evolution of stellar-mass binary PBHs was further elaborated in \cite{1998PhRvD..58f3003I}. More detailed studies of the formation and evolution of binary PBHs have been performed after the LIGO discoveries \cite{2016PhRvL.117f1101S,2016arXiv160404932E,2017PhRvD..96l3523A,2019JCAP...02..018R}. These and many other studies were focused on the explanation of the observed binary BH merging rate inferred from LIGO observations, which is manifestly dependent on the PBH abundance (the parameter $f_\mathrm{PBH}$). 

In this paper, we construct the expected cumulative distribution of the chirp masses $\Mch$ of coalescing primordial binary BHs with a log-normal mass spectrum
of the components taking into account the current detection limits of LIGO/Virgo GW interferometers. The cumulative distribution $F(<\Mch)$ has an advantage that it does not depend sensitively on the uncertain fraction $f_\mathrm{PBH}$ of PBH in dark matter. We compare the model distributions with the one constructed using the published O1-O2 LIGO/Virgo data \citep{2019PhRvX...9c1040A} complemented with independent searches for binary BHs reported by \cite{2020ApJ...891..123N} and estimations of $\Mch$ as inferred from the public LIGO/Virgo O3 data (https://gracedb.ligo.org/ superevents/public/O3/).
We also compare model astrophysical coalescing binary BH distributions calculated in \cite{2019MNRAS.483.3288P}. 
We find that the cumulative distribution $F(<\Mch)$ as inferred from the reported observations can be much better
described by the PBH coalescences with log-normal mass distribution with parameters $M_0\simeq 15-17 M_\odot$ and $\gamma\simeq 0.8-1$ than by astrophysical models considered.

The structure of the paper is as follows.
In Section \ref{s:form}, we briefly remind the reader the model of the PBH formation with log-normal mass spectrum and its features.
In Section \ref{s:model}, we describe the adopted model of binary PBH coalescences. Based on this model, we calculate the relative number of PBH detections with a given chirp mass $\Mch$ by a GW detector with the sensitivity as in O3 LIGO/Virgo that determines the detection limit of a source with given $\Mch$ at the fiducial signal-to-noise ratio (SNR) level $\rho=8$.
In Section \ref{s:EDF}, we describe the construction of the 
empirical chirp mass distribution $P(\Mch)$ and cumulative function $F(<\Mch)$, which are compared with our model distributions $P_\mathrm{PBH}(\Mch)$ and 
$F_\mathrm{PBH}(<\Mch)$. The statistical criteria we use in our analysis are presented in Section \ref{s:statistics}. Section \ref{s:PBHres} describes the results of comparison of the PBH model with empirical distribution function (EDF) constructed in Section \ref{s:EDF}. The comparison of some astrophysical models of binary BH coalescences with EDF is presented in Section \ref{s:astrores}. 
We discuss our results in Section \ref{s:discussion}. In Section \ref{s:conclusion}, we formulate our main findings.

\section{A scenario of PBH formation  with log-normal mass spectrum}
\label{s:form}

It is possible that a significant part (if not all) of black holes in the universe are primordial, for a review 
see \citep{2018PhyU...61..115D}. 
Various astronomical observations have been found to be in reasonable agreement
with a mechanism of PBH formation with extended mass spectrum from a fraction of solar mass up to
billion solar masses which was suggested 
in \citep{1993PhRvD..47.4244D,2009NuPhB.807..229D}.
According to the model elaborated in these papers, 
the PBH mass spectrum at the moment of creation has a log-normal form:
\be
F(M)\equiv \frac{dN}{dM} = \mu^2 \exp \left[- \gamma \ln^2\left( \frac{M}{M_0} \right) \right]  ,
 \label{e:dN-dM}
\ee
where $\gamma$ is a dimensionless constant and parameters $\mu$ and $M_0$ have dimension of mass or, what is the same, 
of inverse length (here the natural system of units with $c=k=\hbar =1$ is used). Probably, the log-normal spectrum is a general feature 
of inflationary production of PBHs or, to be more precise, is a consequence of the creation of appropriate conditions for the PBH formation during the inflationary cosmological stage, while the PBHs themselves might be formed long after the inflation had terminated. 

According to the model of ref.~\citep{1993PhRvD..47.4244D,2009NuPhB.807..229D},
at inflationary stage conditions for a very efficient baryogenesis were created leading to the formation of
astrophysically large bubbles with a very high baryonic number density, much larger than the conventional one,
$\eta=n_B /n_\gamma \sim 10^{-9}$. It can be shown that these bubbles have log-normal distribution over their size. Since quarks were massless in the very early universe, the bubbles with high baryonic number density
had the same energy density as the cosmological background. Therefore, their density contrast was virtually zero.
The situation changed after the QCD phase transition at $T\approx 100$ MeV, when massless quarks turned into massive hadrons (protons and neutrons). The large density contrast which arose 
at that moment led to the formation of PBHs with the log-normal mass spectrum (\ref{e:dN-dM}).
As argued in ref.~\cite{2020arXiv200411669D}, such a scenario implies that the value of $M_0$ should be around 10 solar masses.

Despite this indication, in what follows we will assume that the values of the parameters $\mu$, $\gamma$, and $M_0$ are unknown and will try to estimate $\gamma$, and $M_0$ independently 
from the analysis of available astronomical observations.
This task is highly non-trivial because the original  mass spectrum of PBHs was surely distorted in the
process of subsequent BH evolution
through the matter accretion and possible coalescences (see, e.g., \citep{2020arXiv200302778D,2020arXiv200312589D} and references therein).

The problem of the parameter estimation of the PBH log-normal spectrum
was addressed by us earlier \citep{2015PhRvD..92b3516B,2016JCAP...11..036B}
where we concluded that $\gamma \approx 0.5$ and $M_0 \approx M_\odot$. Under these assumption
it was possible to obtain a reasonable fit to the density of MACHOs (MAssive Compact Halo Objects)  and the amount of BH binaries
registered by LIGO/Virgo observations. However, it became clear from arguments presented in \cite{2020arXiv200411669D} that it impossible to reliably estimate the abundance of PBHs formed by the mechanism suggested in \cite{1993PhRvD..47.4244D} with masses much 
smaller than 10 $M_\odot$ .

Recently, the parameters of distribution (\ref{e:dN-dM}) were estimated in ref. \cite{2019arXiv190510972D}
based on the observed space density of supermassive BHs and the mass spectrum of BHs in the
Galaxy. The conclusion was $\gamma \approx 0.5$ and $M_0 \approx 8 M_\odot$.
This result depends upon the assumption on the evolution of masses of very heavy BHs with
$M>10^3 M_\odot$ due to accretion, which is not well determined, and should be taken with caution.
With these PBH mass spectrum parameters, the predicted number density of MACHOs (if they are PBHs) happened to be 3-4 orders
of magnitude lower than the observed one.
In ref. \cite{2019arXiv190510972D}, possible
ways out of this conundrum were proposed.
However, as follows from ref. \cite{2020arXiv200411669D},
if MACHOs have a primordial origin, they could be rather compact stellar-like objects with solar or subsolar masses formed by the same mechanism as PBHs but
not PBHs because they would have been deep inside the cosmological horizon at the QCD phase transition. Therefore, the
estimates based on the log-normal mass spectrum
down to a fraction of the solar mass are inapplicable.

\section{The model of binary PBH coalescences}
\label{s:model}

The probability that a pair of black holes with unequal masses $M_1$ and $M_2=qM_1$, $q\le 1$, coalesce within the time interval \textit{(t,t+dt)} can be written as  \cite{1997ApJ...487L.139N,1998PhRvD..58f3003I} (see also \cite{2016arXiv160404932E,2019JCAP...02..018R})
\begin{equation}
\label{eq:f}
f(t)dt\propto \int_0^\infty \int_0^\infty \int_0^\infty 
\left(\frac{t}{\tilde t}\right)^{3/37}
\frac{dt}{t} \ F(M_1) F(M_2) F(M_3) dM_1dM_2dM_3 \,.
\end{equation}
In this model, the formation of a binary BH with masses $M_1, M_2$ is assisted by the collision with third BH $M_3$. 
$F(M)$ is the BH mass function which we assume to be of log-normal form (\ref{e:dN-dM}) that follows from the physical model suggested by \cite{1993PhRvD..47.4244D}, Eq. (\ref{e:dN-dM}).

Following  \citep{1998PhRvD..58f3003I}, we define the normalized time as
\begin{equation}
\label{eq:tildet}
\tilde t = (\eta \tilde \beta)^7 \left(\frac{\xi \tilde \alpha \bar x}{a_0} \right)^4 \ t_0\,.
\end{equation}
Here $\bar x$ is the mean separation of black holes  with mass $M$ at the time of matter-radiation equality, 
$t_0= 13.8\times 10^{9}$ years and $a_0$ is the semimajor axis of a binary with circular orbit which coalesces due to GW emission in $t_0$.
The fractional masses are defined as 
\begin{equation}
\label{eq:eta}
\eta = \frac{2 M_3}{M_1 + M_2}, \quad \xi = \frac{2 \overline M_{BH} }{M_1 + M_2}
\end{equation}
with the mean BH mass
\begin{equation}
\label{eq:meanM}
\overline M_{BH} = \int_0^\infty M F(M)dM 
\end{equation}
$\tilde \alpha,\ \tilde \beta$ are numerical constants of order $O(1)$ which we will set equal to unity. Their numerical value affects 
the volume rate of binary BH coalescences \cite{1998PhRvD..58f3003I,2016PhRvL.117f1101S,2016arXiv160404932E} 
but virtually does not change the chirp mas distribution of coalescing BH binaries studied here.

Denoting the comoving binary BH number density as $n_{BH}$, the binary BH  coalescence rate per unit comoving volume per year reads\footnote{The suppression factor $C(f_{PBH})<1$ that appears in this expression in more advanced models, e.g. \cite{2017PhRvD..96l3523A,2019JCAP...02..018R}, does not affect the chirp mass distribution; see Section \ref{s:discussion}. }:
\begin{equation}
\label{eq:R}
\mathcal{R} = n_{BH} f(t)\,.
\end{equation}
 According to the model of \cite{1993PhRvD..47.4244D}, 
the characteristic time of the first PBH formation is of the order of the phase transition time when they 
happened to enter inside the horizon, which is much shorter than the Hubble time. The time of  binary formation may be rather extended depending upon the chaotic initial separation of PBHs  and the dynamics of three-body collisions considered in refs. \cite{1998PhRvD..58f3003I,2016arXiv160404932E}. However, for the late time evolution of interest here we will assume that all binary PBHs were formed instantly in the early Universe and no new sources were produced. The assumption that newly formed PBHs are not disrupted before merging  by tidal interactions with other PBHs 
depends on the factor $f_\mathrm{PBH}$ and holds if $f_\mathrm{PBH}\ll 1$ \cite{2019JCAP...02..018R}.

\begin{figure}
    \centering
    \includegraphics[width=0.9\textwidth]{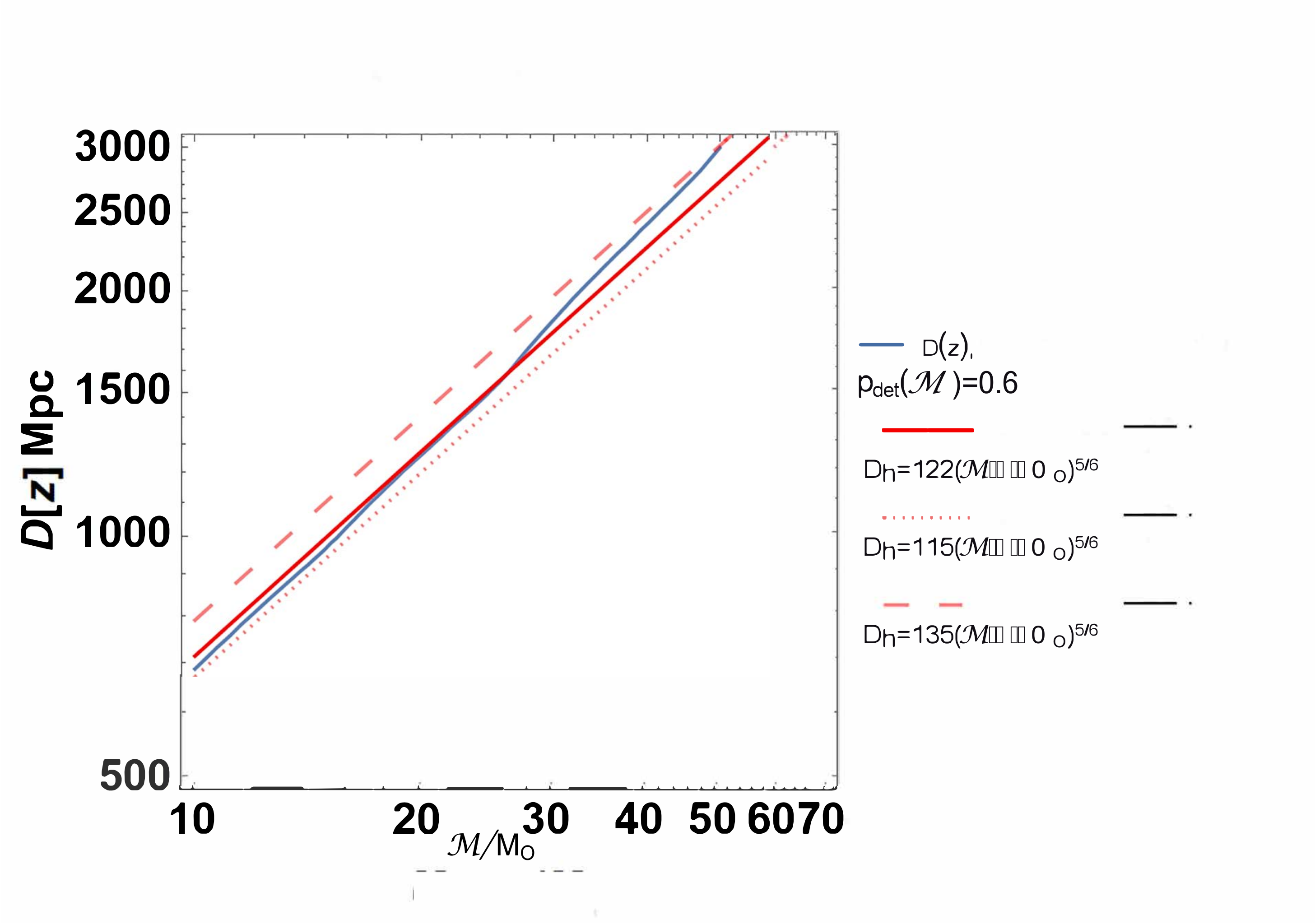}
    \caption{\added{The detection horizon as a function of the binary chirp mass calculated using the criterion $p_\mathrm{det}(M_1,M_2,z)=0.6$ for the detection probability from \cite{2017zndo....889966G} (IMRPhenomD waveforms, the blue curve)
    and the approximation according to Eq. (\ref{eq:Dh}) (the red curve).}}
    \label{f:Dh}
\end{figure}

The detection rate of binary BH mergings per year by a detector with given sensitivity that determines the detection horizon $D_h(\mathcal M)$ is:
\begin{equation}
\label{eq:DR}
\mathcal{DR}(\Mch) =\int\limits_0^{z(D_h(\mathcal M))} \frac{\mathcal R}{1+z'} \ \frac{dV}{dz'} \ dz'
\end{equation}
\begin{equation}
\label{eq:Dh}
D_h(\mathcal M) = 122 \mathrm{Mpc} \left( \frac{\mathcal{M}}{1.2 \ M_\odot} \right)^{5/6}\,.
\end{equation}
For inspiraling binaries, the detection horizon is mostly dependent on the chirp mass $\Mch$ \cite{1993PhRvD..47.2198F,
2010arXiv1003.2481T}. Alternatively, one could use the detection rate as a function of individual masses $M_1$ and $M_2$, then the non-trivial probability detection factor $p_\mathrm{det}(M_1,M_2,z)$ should be used (see, e.g., ref. \cite{2020JCAP...01..031G}). However, in coordinates $\Mch, q=M_2/M_1$, this factor is independent of the mass ratio $q$, and the dependence (\ref{eq:Dh}) is a good approximation for the detection sensitivity for optimally oriented binary with the numerical coefficient 122 Mpc reflecting the sensitivity of real O3 detectors (see Fig. \ref{f:Dh}).

While the coalescence rate of binary PBHs manifestly depends on the assumed fraction of PBHs $f_\mathrm{PBH}$, it should be clear that the cumulative distribution of the PBH coalescences by their chirp mass $\Mch$, $F_\mathrm{PBH}(<\Mch)$, i.e. the fraction of detected sources with $\Mch$ smaller than a given one, is independent of this uncertain parameter (see also Section \ref{s:discussion} below). This fraction is calculated as
\beq{e:x}
F_\mathrm{PBH}(<\Mch)=\int\limits_0^\Mch P_\mathrm{PBH}(\Mch') d\Mch'=\frac{\int\limits_0^\Mch \mathcal{DR}(\Mch')d\Mch'}{\int\limits_0^\infty \mathcal{DR}(\Mch')d\Mch'}
\eeq

\section{Observational sample}
\label{s:EDF}

To compare the model distribution $F_\mathrm{PBH}(<\Mch)$ (or its PDF $P_\mathrm{PBH}(\Mch)$ with observations, we need to construct a sample of the observed BH+BH chirp masses. To this goal, we have used BH+BH chirp masses from the following sources.
\begin{enumerate}
    \item BHBHs from the LIGO/Virgo GWTC-1 catalog \citep{2019PhRvX...9c1040A};
    \item BHBHs found in O1-O2 LIGO/Virgo runs by independent searches reported in \citep{2020ApJ...891..123N};
    \item One BHBH GW190412 from LIGO/Virgo O3 run \citep{2020arXiv200408342T}.\footnote{Adding new O3 BH-BH binary GW190521 \cite{2020arXiv200901190T} with $\Mch\approx 65\,M_\odot$ does not affect the results of our analysis.}
\end{enumerate}

Clearly, this sample is not complete and will be extended by new sources from O3. 
Nevertheless, we attempted to complement it by estimates of $\Mch$ derived from public O3 data. To do this, we assumed that the reported O3 sources are detected at the minimum accepted signal-to-noise ratio (SNR) $\rho_\mathrm{min}=8$. Then the chirp mass of the detected source can be estimated using the O3 LIGO/Virgo detection horizon, $D_h=122\,\mathrm{Mpc}$ for the canonical binary neutron star chirp mass $\Mch_\mathrm{NSNS}=1.22\, M_\odot$ and the reported estimate of the luminosity distance to the source $D_l$, from the relation $D_h(\Mch)=D_l$ (see Eq. (\ref{eq:Dh})). 

Of course, this is a lower limit to the chirp mass because actual SNR for each source can be higher than the threshold value SNR=8, but we assume that on average, for $50$ O3 sources, their real chirp mass distribution will be not too different from our estimate. 
In this way we can evaluate the detected chirp mass of each source $\Mch_\mathrm{det}$, which then can be recalculated to the source frame $\Mch_\mathrm{s}=\Mch_\mathrm{det}/(1+z)$, where $z$ is the source redshift derived from the reported photometrical distance $D_l$. We use the standard cosmology with parameters reported by the Planck collaboration \citep{2018arXiv180706209P}. For example, our estimate of the chirp mass of GW190412 derived in this way is $\Mch_\mathrm{det}\approx 12.1 M_\odot$, $\Mch_\mathrm{s}\simeq 10.4 M_\odot$, while the LIGO/Virgo reported value is $\Mch_\mathrm{s}\approx 13.2 M_\odot$ \citep{2020arXiv200408342T}, the difference being mainly due to its actual SNR $\rho\approx 19$. 

By assuming the minimum adopted SNR, our procedure produces the lower value of the actual chirp masses. To take into account this effect, the most reliable sources with known SNR from \cite{2019PhRvX...9c1040A} and \cite{2020ApJ...891..123N} (27 sources) were ascribed weights calculated as $w_i=\rho_i/\rho_{\min}$, where $\rho_i$ is the SNR of the i-th detected source calculated as the average between $\rho_H$ and $\rho_L$ from Table 3 in \citep{2020ApJ...891..123N},  $\rho_{\min}\approx 6$ is the minimum value of this sample. New O3 sources (but GW190412) with unknown SNR (49 sources) were all ascribed $\rho=8$. To take into account different sensitivity in O1-O2 and O3 LIGO/Virgo runs, the weights of all O1-O2 sources were additionally increased by factor $120/90$ roughly corresponding to the increase in the O3 LIGO/Virgo sensitivity relative to O1-O2 runs (i.e., the reported SNR for each of O1-O2 sources was artificially increased by this factor as if they were observed by detectors with O3 LIGO/Virgo sensitivity).

Thus constructed empirical distribution $F(<\Mch)$ for the entire sample of 77 sources is shown in Figs. \ref{f:KS_EDF}, \ref{f:VdW_EDF} and \ref{f:astro_EDF} by the blue step line. 

\section{Statistical tests}
\label{s:statistics}

Consider the observed distribution over the chirp masses of merging double BHs  obtained by the gravitational-wave interferometers LIGO/Virgo $P(\Mch)$. The empirical distribution function $P(\Mch)$ can be compared with model distributions
$P_\mathrm{PBH}(\Mch)$ using two statistical tests: modified Kolmogorov-Smirnov (KS) and Van der Waerden test (VdW). 
The model distribution $P_\mathrm{PBH}(\Mch)$ is the normalized number of coalescing binary BHs calculated in the model 
\cite{1997ApJ...487L.139N,1998PhRvD..58f3003I} for the log-normal PBH mass spectrum (\ref{e:dN-dM}) with parameters $M_0$ and $\gamma$. 

In order to compare the model distribution calculated for various parameters $7 \le M_0/M_\odot \le 19$, $0.5 \le \gamma \le 1.9$ with the empirical data, we construct theoretical cumulative distributions $F_\mathrm{PBH}(<\Mch)$ with the same number of bins as for EDF  $F(<\Mch)$. 

The null hypothesis reads:
$H_0=\{$\textit{Two samples $F_\mathrm{PBH}(<\Mch)$  and  $F(<\Mch)$ are arbitrary taken from one distribution (are equal)}$\}$. 

To check this hypothesis we use KS and VdW statistical tests.
\begin{enumerate}
\item 
The error of the classical  Kolmogorov-Smirnov criterion for a sample size $n>60$ does not exceed 0.8\% \cite{KendallStuart1973}.
However, according to recent studies, the convergence of the test is relatively slow (for a sample of size $n = 100$, the maximum error is still about 2.6\% 
\cite{Vrbik2018}).
In our analysis, the total sample size is $n = 27+50=77$. Taking into account the 
recommendation from recent statistical research, we use the modified KS criterion, which is effective when comparing cumulative distributions of small samples. The method is based on finding the maximum difference between elements from two samples with the same  element numbers. The corresponding statistic uses the following formula \citep{Vrbik}:
\begin{equation}
\label{eq:KS}
\sup(|x_1(i)-x_2(i)|)+\dfrac{1}{\sqrt{6}n}+\dfrac{\sup(|x_1(i)-x_2(i)|)-1}{4n},
\end{equation}
where $x_1(i)$ is the i-th element from the 1st sample, and $x_2 (i)$ is the i-th element from the 2nd sample. This statistic makes it possible to conclude whether the null hypothesis is accepted or rejected.

\item 
To accept the null hypothesis 
of equality of two samples, it is necessary to check the equality of both mean and variance of two samples. The equality of the mean values of two distributions is checked using the Van der Waerden test. This non-parametric (distribution-free) test is effectively applied to small samples that contain duplicated elements. Our empirical sample (EDF $F(<\Mch)$) contains such elements introduced in order to account for data weights according to the accuracy (the signal-to-noise ratio) of the observations (see Section \ref{s:EDF}).  

The VdW statistic has the form
\begin{equation}
\label{eq:X}
X=\sum_{i=1}^M u_{{R_i}/{M+N+1}},
\end{equation}
where $N, M$ are the number of elements in the samples, $ u_{{R_i}/{M+N+1}}$ is quantile of the standard normal distribution $N(0,1)$. To calculate the quantiles of $ u_{{R_i}/{M+N+1}}$, the following approximation can be applied:
\begin{equation}
\label{eq:uR}
u_{{R_i}/{M+N+1}}=4,91 \left[ \left( \frac{R_i}{M+N+1} \right)^{0,14}-\left( 1 - \frac{R_i}{M+N+1} \right)^{0,14} \right]
\end{equation}
If $|X|<x_\delta$, then with the confidence probability $\delta$ ($x_\delta$ is the critical value of the VdW statistic), the hypothesis that both samples have the same mean is accepted.

To compare the variance of two samples, the statistics 
\beq{e:Z}
Z = \dfrac{ \bigg|R_2 - \dfrac{M(N+M+1)}{2}\bigg| - \dfrac{1}{2} }{ \sqrt{\dfrac{N(N+M+1)}{12}} }.
\eeq
is used. 
The statistics is constructed as follows. A joint sample
of size $N+M$ is arranged in the increasing order 
and individual elements from each sample are numbered. Ranks are assigned according to the following rule: the smallest element is assigned rank 1, two maximum elements are assigned rank 2 and 3, ranks 4 and 5 are assigned to the next minimal elements, etc. Thus the rank scheme is: (1, 4, 5, 8, 9, \ldots , 7, 6, 3, 2).
Each of the coincident elements is assigned the rank equal to the mean value. In formula (\ref{e:Z}) $R_2$ is the sum of ranks for the smaller sample ($M \le N$). 
If the sample variances are equal, the $Z$ statistics is distributed close to $N(0,1)$. 
The hypothesis that the variances are equal is rejected if
\beq{}
|z|>u_{1-\alpha/2},\quad \alpha=1-\delta,\quad \delta=95\%.
\eeq
\end{enumerate}
These criteria were applied for different values of constants $M_0$ and $\gamma$ in the next Section.

\section{Comparison of model PBH chirp mass distributions with EDF}
\label{s:PBHres}

Model distributions $P_\mathrm{PBH}(\Mch)$ and $F_\mathrm{PBH}(<\Mch)$ of coalescing PBH chirp masses calculated for the range of parameters $7\le (M_0/M_\odot)\le 19$ and $0.5\le \gamma\le 1.9$
were compared with the empirical PDF  $P(\Mch)$ and cumulative distribution $F(<\Mch)$ using the statistical KS test (for cumulative distributions),
Van der Waerden test to check the equality of the mean sample values, and $Z$-statistics for comparison of the sample variances. 

\begin{figure}
    \centering
    \includegraphics[width=\textwidth]{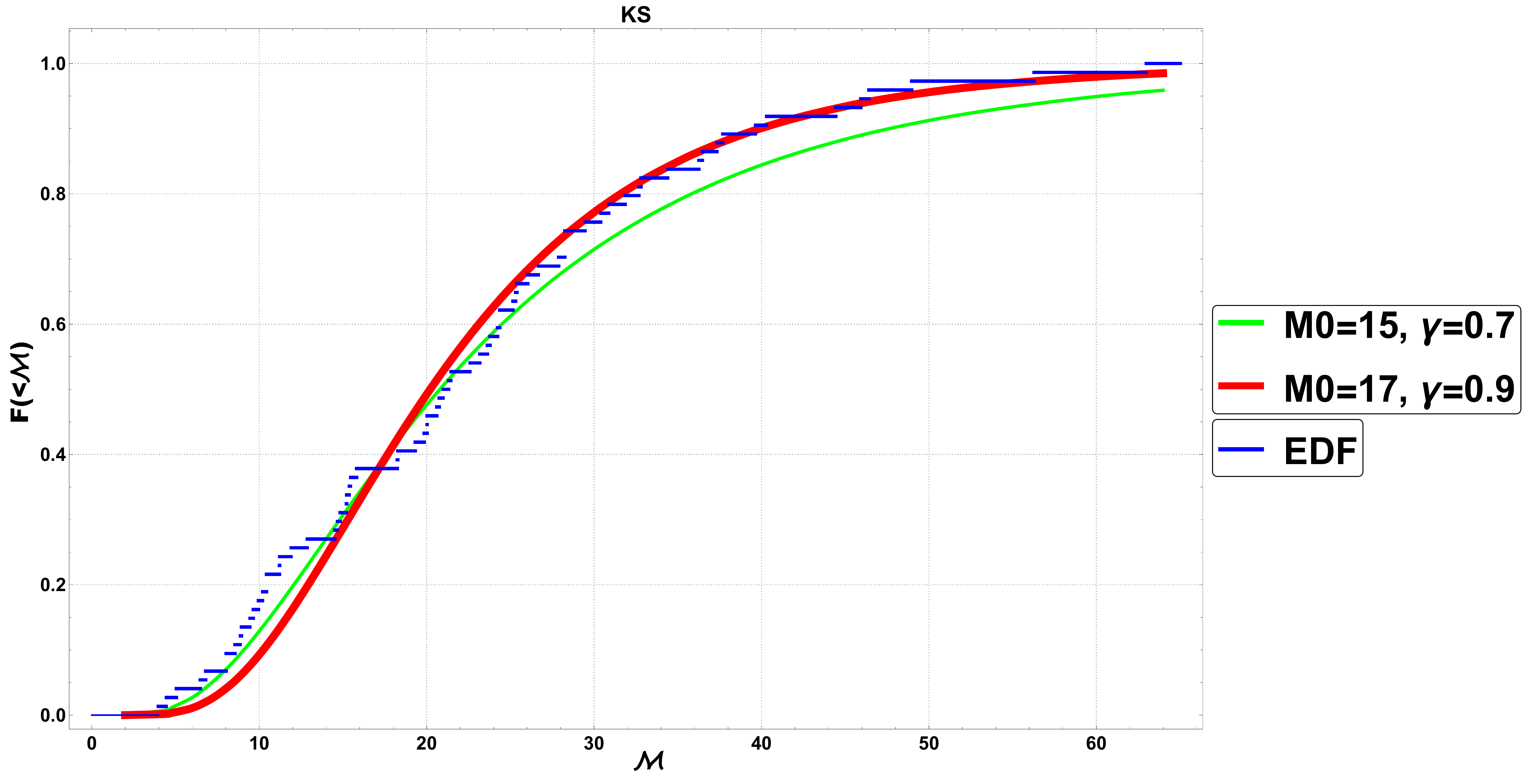}
    \caption{Model distribution $F_\mathrm{PBH}(<\Mch)$ with parameters $M_0,\gamma$ for two best KS-tests from Table \ref{t:PBHKS}: $M_0=17M_\odot, \gamma=0.9$ (red line), $M_0=15M_\odot, \gamma=0.7$ (green line). The empirical distribution $F(<\Mch)$ is shown by the blue step curve.}
    \label{f:KS_EDF}
\end{figure}

Table \ref{t:PBHKS} lists the KS-test for the set of PBH cumulative functions $F_\mathrm{PBH}(<\Mch)$. Models for which null hypothesis cannot be rejected at 95\% level are marked in green. The best model corresponds to $M_0=15 M_\odot$ and $\gamma=0.7$ (in red in Table \ref{t:PBHKS}). This model cumulative distribution and that having the second best KS value from Table \ref{t:PBHKS} ($M_0=17 M_\odot$ and $\gamma=0.9$) are plotted in Fig. \ref{f:KS_EDF} in green and red lines, respectively.

\begin{figure}
    \centering
    \includegraphics[width=\textwidth]{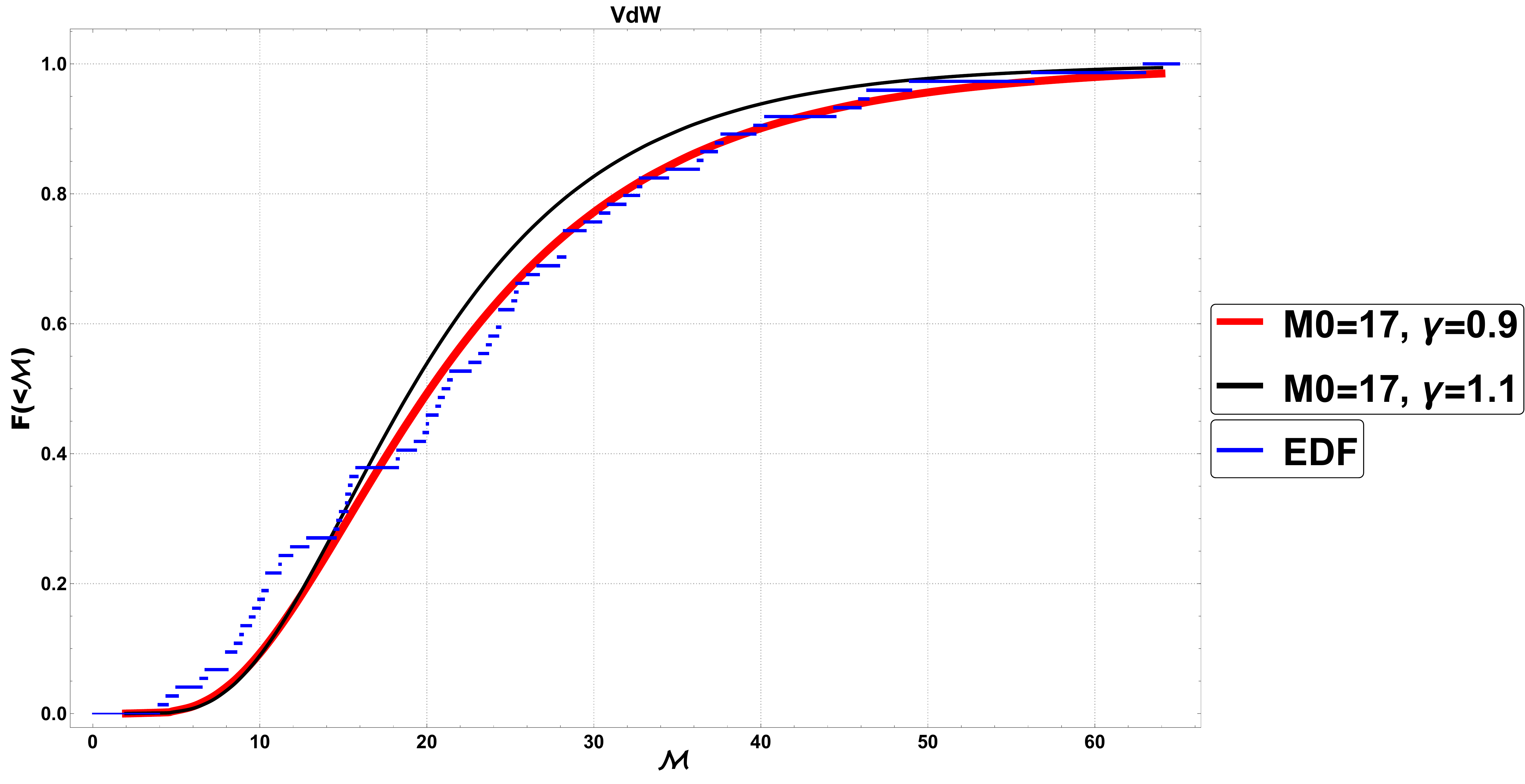}
    \caption{Model distribution $F_\mathrm{PBH}(<\Mch)$ with parameters $M_0$ and $\gamma$ for two best VdW-tests from Table \ref{t:PBHVdW}: $M_0=17M_\odot, \gamma=0.9$ (red line), $M_0=17M_\odot, \gamma=1.1$ (black line). The empirical distribution $F(<\Mch)$ is shown by the blue step curve.}
    \label{f:VdW_EDF}
\end{figure}

The results of comparison of the mean values and variances of the empirical PDF $P(\Mch)$ and model PDFs $P_\mathrm{PBH}(\Mch)$ by VdW test and $Z$ statistics are presented in Tables \ref{t:PBHVdW} and \ref{t:PBHVdW_var}, respectively. The models for which null hypothesis cannot be rejected at 95\% level are marked in green, the best model is shown in red. Both VdW test and $Z$ statistics single out the same model with $M_0=17 M_\odot$ and $\gamma=0.9$.
Cumulative distributions $F_\mathrm{PBH}$ for this and the second best model according to VdW test ($M_0=17 M_\odot$, $\gamma=1.1$) are plotted in Fig. \ref{f:VdW_EDF} in red and black lines, respectively.

Thus, the statistical criteria we applied suggest that the chirp mass distribution calculated for the PBH log-normal mass spectrum with $M_0=17 M_\odot$ and $\gamma=0.9$ does not contradict the empirical PDF $P(\Mch)$. The PBH central mass $M_0\sim 17 M_\odot$ may be in line with lower temperature of the QCD phase transition at non-zero chemical potential.
\begin{table}[ht]
\caption{\label{t:PBHKS}Comparison of PBH models $F_\mathrm{PBH}(<\Mch)$ with EDF $F(<\Mch)$ using modified Kolmogorov-Smirnov test. In green shown are models for which null hypothesis cannot be rejected at the 95\% level (KS$<1.36$). The best model is shown in red.}
\begin{center}
\begin{tabular}{ | c | l  l  l  l  l  l  l |}
\hline
\backslashbox{$\gamma$}{$M_0/M_\odot$}
& 7 & 9 & 11 & 13 & 15 & 17 & 19 \\ 
\hline
\hline
0.5 & 3.20 &	2.41 & 1.66 & \textcolor{green}{0.95} &	\textcolor{green}{1.31} &	1.85 &	2.34 \\ 
0.7 & 4.54 &	3.52 &	2.64 &	1.77 &	\textcolor{red}{0.64} &	\textcolor{green}{0.93} &	\textcolor{green}{1.32} \\ 
0.9 & 5.08 &	4.04 &	3.15 &	2.42 &	\textcolor{green}{1.29} &	\textcolor{green}{0.89} &	\textcolor{green}{1.21} \\ 
1.1 & 5.45 &	4.40 &	3.68 &	2.90 &	1.88 &	\textcolor{green}{1.11} &	\textcolor{green}{1.24} \\
1.3 & 5.67 &	4.78 &	3.98 &	2.68 &	2.24 &	1.44 &	\textcolor{green}{1.26} \\ 
1.5 & 5.83 &	5.05 &	4.24 &	3.52	& 2.66 &	1.72 &	\textcolor{green}{1.30} \\ 
1.7 & 5.95 &	5.28 &	4.42 &	3.74 &	2.89 &	1.94 &	\textcolor{green}{1.33} \\ 
1.9 & 6.05 &	5.44 &	4.57 &	3.94 &	3.11 &	2.12 &	1.37 \\
\hline
\end{tabular}
\end{center}
\end{table} 

\begin{table}[ht]
\caption{\label{t:PBHVdW}Comparison of the mean of  model PDF $P_\mathrm{PBH}(\Mch)$ with that of the empirical distribution $P(\Mch)$ using Van der Waerden test (VdW). In green shown are models for which null hypothesis that both samples have the same mean  cannot be rejected at 95\% level (VdW$<1.96$). The best model is shown in red.}
\begin{center}
\begin{tabular}{|  c | l  l  l  l  l  l  l  |}
\hline
\backslashbox{$\gamma$}{$M_0/M_\odot$}
& 7 & 9 & 11 & 13 & 15 & 17 & 19 \\ 
 \hline
 \hline
0.5 & 4.01&	2.57&	\textcolor{green}{1.25}&	\textcolor{green}{0.04}&	\textcolor{green}{1.61}&	2.66&	3.47 \\ 
0.7 & 5.91&	4.40&	2.81&	\textcolor{green}{1.36}&\textcolor{green}{	0.64}&	\textcolor{green}{1.43}&	2.17 \\ 
0.9 & 6.47&	5.05&	3.52&	2.33&	\textcolor{green}{0.66}&	\textcolor{red}{0.26}&	\textcolor{green}{1.28} \\ 
1.1 & 6.81&	5.51&	4.17&	2.93&	\textcolor{green}{1.45}&	\textcolor{green}{0.36}&	\textcolor{green}{0.80} \\ 
1.3 & 6.99&	5.85	&4.49&	2.60	&\textcolor{green}{1.90}&	\textcolor{green}{0.73}&	\textcolor{green}{0.33} \\ 
1.5 & 7.11&	6.03&	4.75&	3.57&	2.37&	\textcolor{green}{1.07}&	\textcolor{green}{0.03} \\ 
1.7 & 7.23&	6.19&	4.93&	3.75&	2.57&	\textcolor{green}{1.29}&	\textcolor{green}{0.16} \\ 
1.9 & 7.31&	6.27&	5.07	&3.91&	2.70&	\textcolor{green}{1.39}&	\textcolor{green}{0.33} \\
\hline
\end{tabular}
\end{center}
\end{table} 
\begin{table}[ht]
\caption{\label{t:PBHVdW_var}Comparison of the variance of the PBH distribution $P_\mathrm{PBH}(\Mch)$ with that of EDF $P(\Mch)$ using $Z$ statistics (\ref{e:Z}). In green shown are models for which null hypothesis that both samples have the same variance cannot be rejected at 95\% level. The best model is shown in red.}
\begin{center}
\begin{tabular}{|  c | l  l  l  l  l  l  l  |}
\hline
\backslashbox{$\gamma$}{$M_0/M_\odot$}
& 7 & 9 & 11 & 13 & 15 & 17 & 19 \\ 
 \hline
 \hline
0.5 &  36.0 & 23.7& 12.7& 2.11 & 11.3 &	20.4& 27.7 \\ 
0.7 &  52.4 & 38.7& 25.4& 13.4 & 3.02 &	9.75& 16.1 \\
0.9 &  57.9 & 44.8& 31.3& 21.3 & 7.41 &	\textcolor{red}{0.23}& 8.84 \\
1.1 &  61.3 & 49.2& 37.4& 26.6 & 14.1 &	4.79& 4.62 \\
1.3 &  63.3 & 52.6& 40.5& 23.7 & 17.9 &	7.83& \textcolor{green}{1.31} \\
1.5 &  64.7 & 54.6& 43.0& 32.5 & 21.9 &	10.5& \textcolor{green}{1.13} \\
1.7 &  65.9 & 56.3& 44.9& 34.3 & 24.1 &	12.5& 3.05 \\
1.9 &  66.8 & 57.3& 46.4& 35.9 & 25.6 &	13.9& 4.48 \\
\hline
\end{tabular}
\end{center}
\end{table}

\section{Astrophysical models for binary BH chirp mass}
\label{s:astrores}

For completeness of our analysis, we also constructed the chirp mass distribution $F_\mathrm{aph}(<\Mch)$ of astrophysical coalescing binary BHs calculated in two specific BH formation models discussed earlier in 
\citep{2019MNRAS.483.3288P}. In the first model (below referred to as 'CO'), we have assumed that BHs result from the direct collapse of the C-O core of massive rotating stars, with $M_{BH}=0.9M_{CO}$ and without additional fallback from the stellar envelope. In the second model (dubbed 'BH' below), the BH mass is calculated according to Ref.  \citep{2012ApJ...749...91F}.
The effective spins of binary black holes in this study are calculated with a model account of tidal interactions and effective core-envelope coupling.

\begin{figure}
    \centering
    \includegraphics[width=0.495\textwidth]{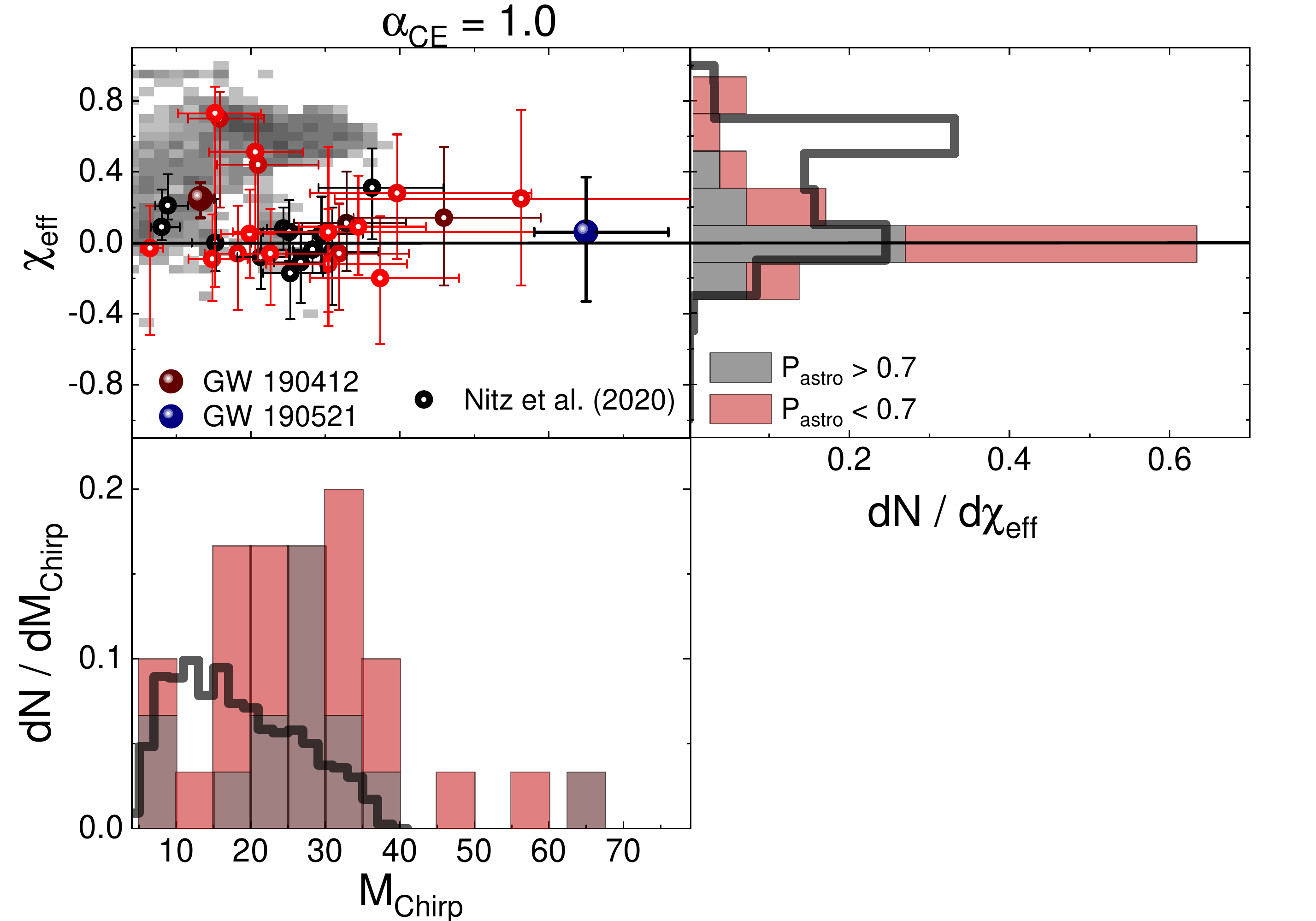}
     \includegraphics[width=0.495\textwidth]{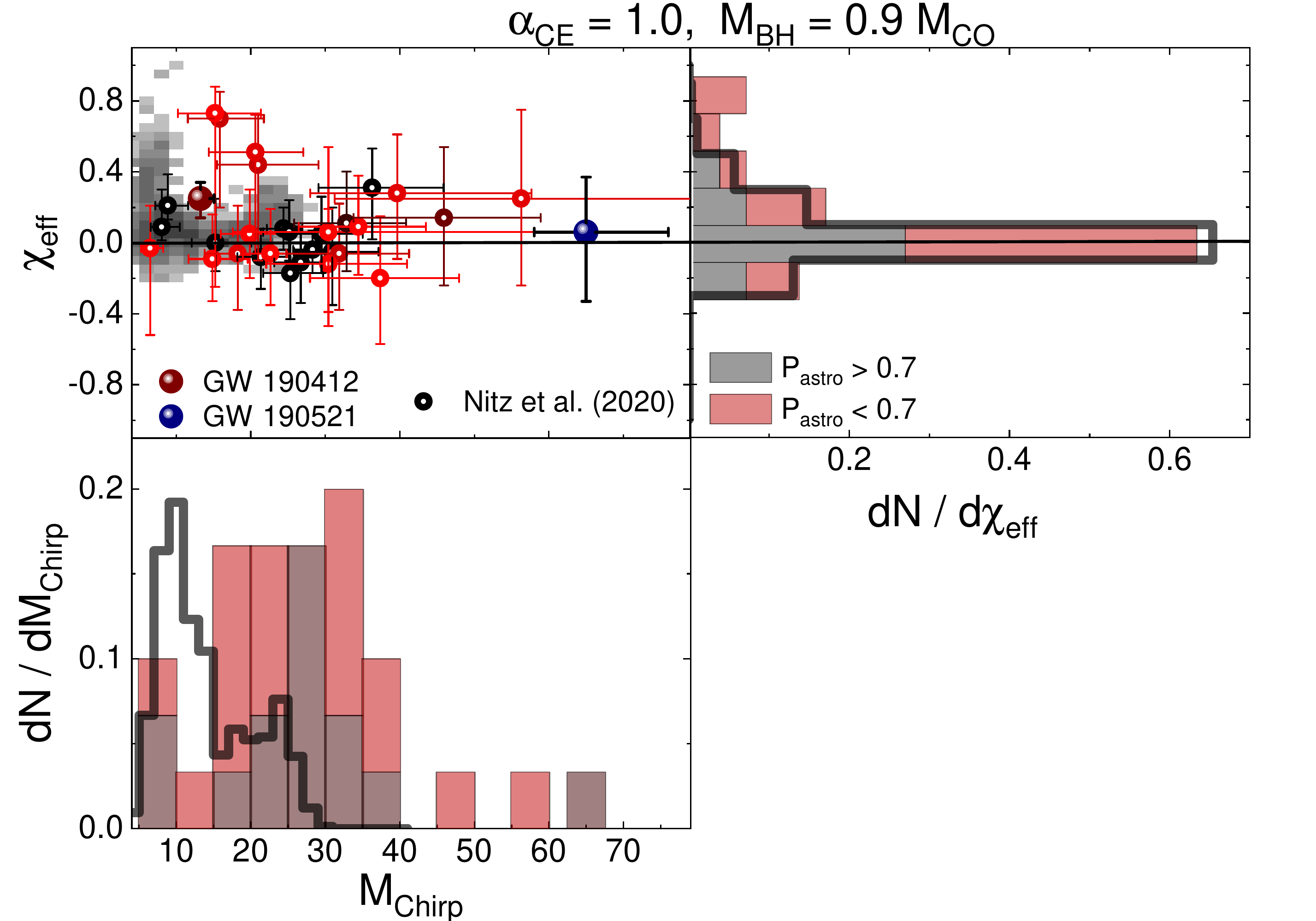}
     \includegraphics[width=0.495\textwidth]{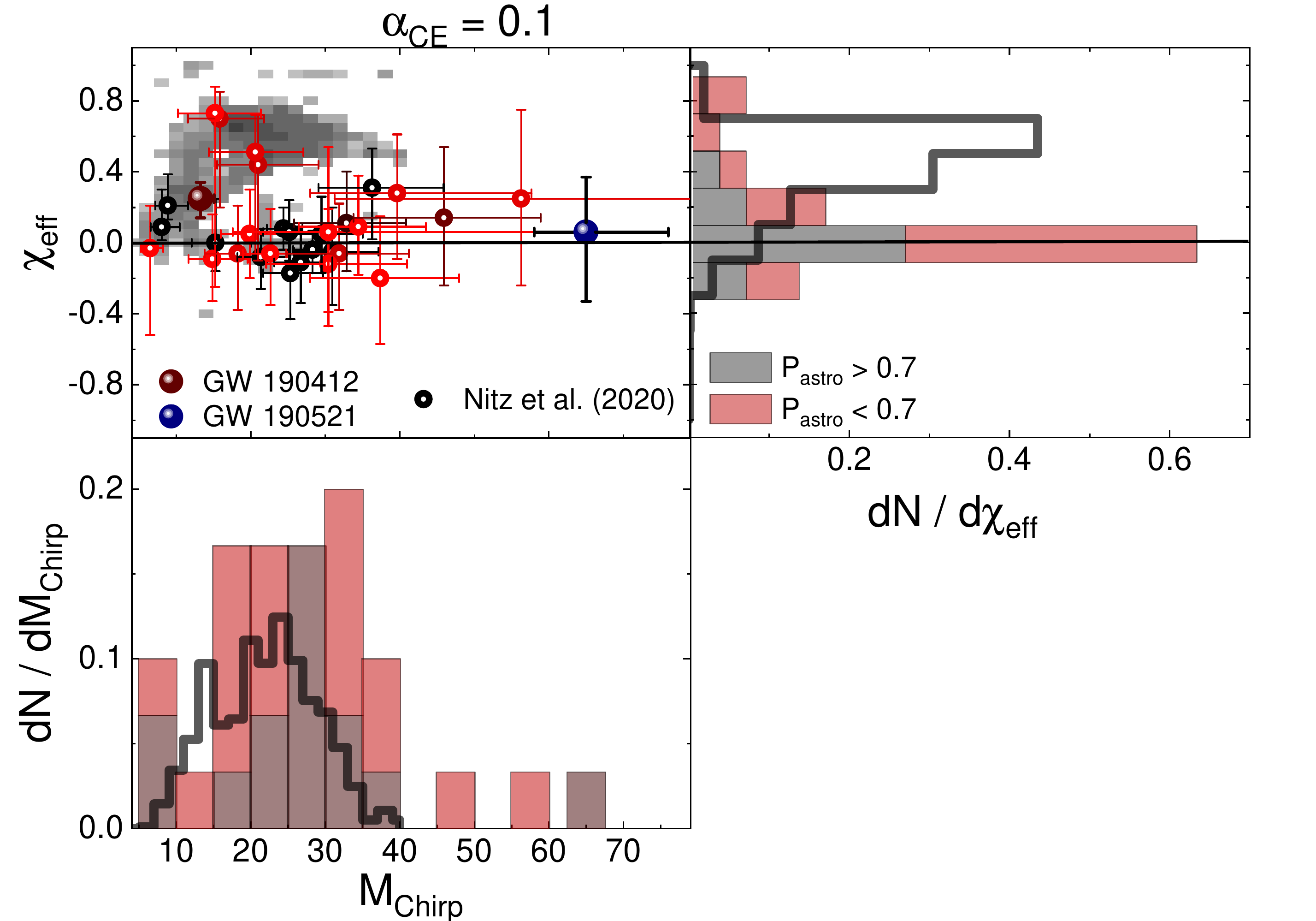}
     \includegraphics[width=0.495\textwidth]{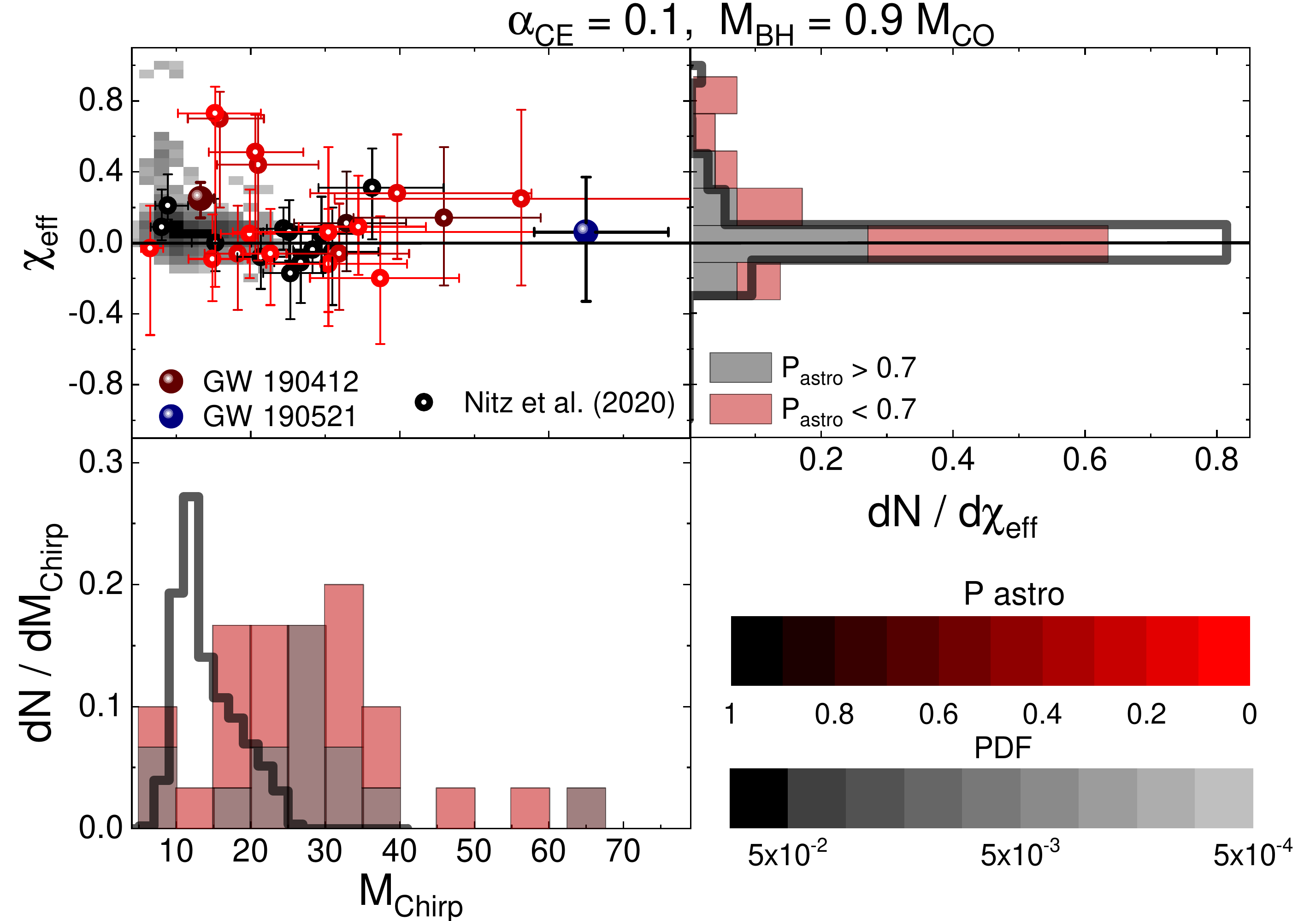}
    \caption{Chirp mass $\Mch$ and effective spin $\chi_\mathrm{eff}$ distribution of coalescing binary BHs that can be detected with the O3 LIGO/Virgo sensitivity for two astrophysical binary BH formation models 'BH' (left panels) and 'CO' (right columns) calculated with account for the star formation rate and stellar metallicity evolution in ref. \citep{2019MNRAS.483.3288P}, for two common envelope efficiency parameters $\alpha_\mathrm{CE}=1$ (upper row) and $\alpha_\mathrm{CE}=0.1$ (bottom row). Red color scale marks the 'astrophysical probability' of sources from ref. \cite{2020ApJ...891..123N}. Grey scale is the PDF of simulations from ref. \citep{2019MNRAS.483.3288P}. See text for more detail. }
    \label{f:astro}
\end{figure}

Fig. \ref{f:astro} in grey color shows distributions of chirp mass $\Mch$ and effective spin $\chi_\mathrm{eff}$ of the population of coalescing binary BHs that can be detected at the O3 LIGO sensitivity (Eq. \ref{eq:Dh}).  The cosmological star formation rate and stellar metallicity evolution is taken into account \citep{2019MNRAS.483.3288P}. As an example, we show the results for two model differing by the common envelope efficiency parameter $\alpha_\mathrm{CE}=1$ (upper row) and $\alpha_\mathrm{CE}=0.1$ (bottom row). The smaller $\alpha_\mathrm{CE}$, the more efficient is the common envelope, i.e. the smaller is the final separation of the binary components after the common envelope stage. The 'BH' model (left column in Fig.\ref{f:astro}) better reproduces the chirp mass distribution (black solid histogram on the bottom left panels) than the 'CO' model (right column in Fig. \ref{f:astro}). Note that in both models large chirp masses cannot be obtained due to the assumed BH 'pair instability gap' \cite{2017ApJ...836..244W}. The 'BH' model enabling the fallback from the rotating stellar envelope shows broader effective spin distribution that the 'CO' model (right top panels). 

In Fig. \ref{f:astro}, in black (more reliable) and red (less reliable) symbols with error bars shown are BHBH sources from Table 3 in \cite{2020ApJ...891..123N}.
The position of GW190412 and GW190521 in this plot is shown by the purple and blue bullets, respectively.

\begin{figure}[ht]
    \centering
    \includegraphics[width=\textwidth]{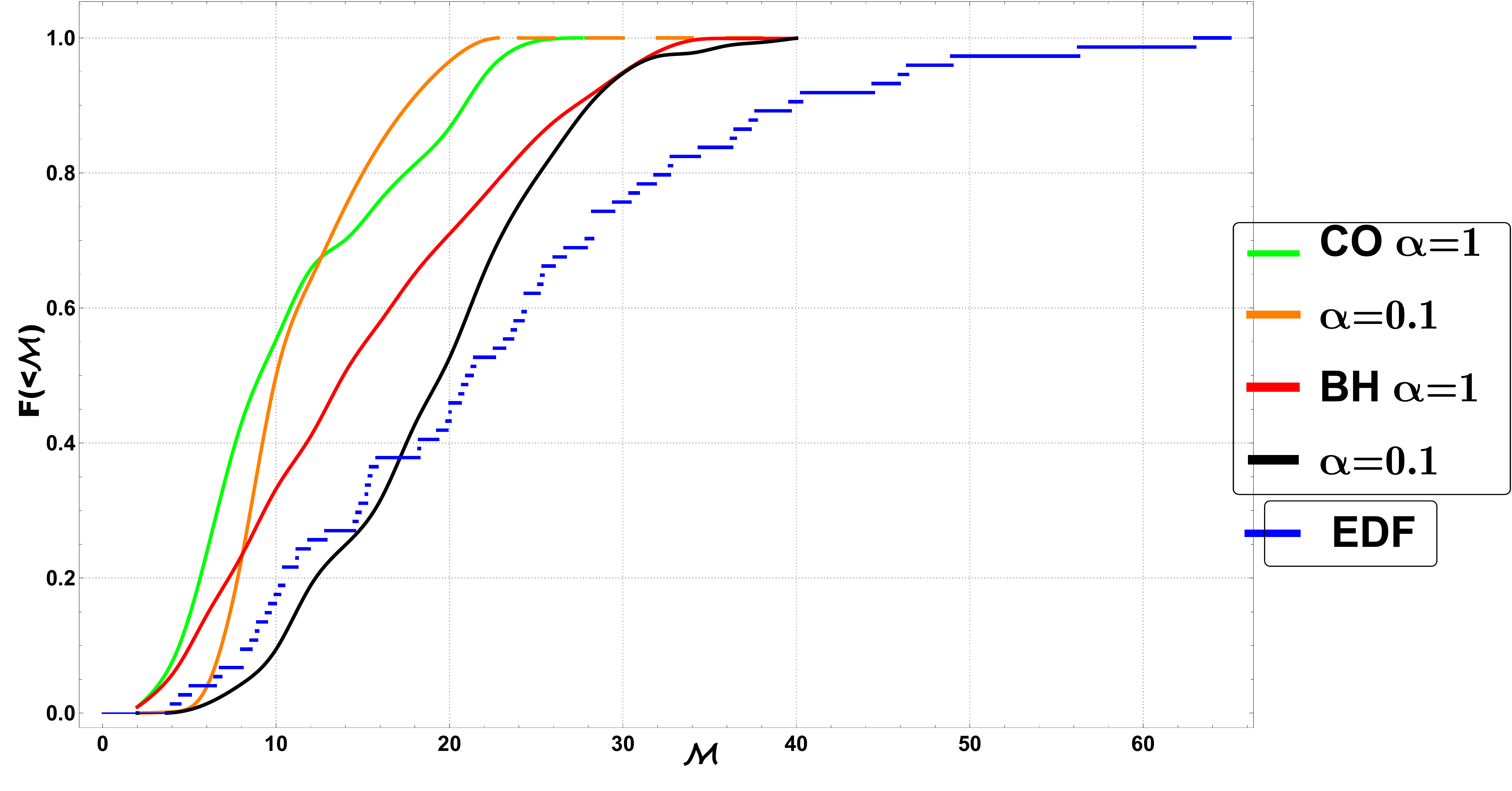}
    \caption{Cumulative distributions $F_\mathrm{aph}(<\Mch)$ for astrophysical models of binary BH coalescences 'CO' and 'BH' (solid curves) for common envelope parameters $\alpha_{\mathrm{CE}}=0, 0.1$. The empirical distribution $F(<\Mch)$ is shown by the blue step curve.}
    \label{f:astro_EDF}
\end{figure}

The cumulative distributions $F_\mathrm{aph}(<\Mch)$ are compared with EDF in Fig.\ref{f:astro_EDF}. The results of KS and VdW tests for models with different common envelope efficiencies $\alpha_\mathrm{CE}$ are listed in Table \ref{t:astro}. According to VdW test, only one model 'BH' with $\alpha_\mathrm{CE}=0.1$ cannot be rejected at the 95\% level. However, it looks much worse than the best PBH model $F_\mathrm{PBH}(<\Mch)$ with $M_0=17 M_\odot$ and $\gamma=0.9$ shown by the red line in Figs. \ref{f:KS_EDF} and \ref{f:VdW_EDF}. 
In Fig. \ref{f:astro}, left bottom panel, the distribution of $\Mch$ and $\chi_\mathrm{eff}$ for the model 'BH' with $\alpha_\mathrm{CE}=0.1$ is presented. It is seen from this Figure that while the chirp mass distribution of 
coalescing binary BH in this model roughly follows $\Mch$
for O1-O2 sources (bottom left frame), as the VdW test for the mean values of the samples suggests, the predicted effective spins are strongly off the observed distribution centered at zero (upper right frame). On the other hand, the 'CO' model (right panels in Fig. \ref{f:astro}) better reproduces the effective spins of the coalescing binary BHs but does not fit their chirp mass distribution for both values of $\alpha_\mathrm{CE}$.

\begin{table}[ht]
\caption{\label{t:astro}
Comparison of astrophysical models $F_\mathrm{aph}(<\Mch)$ with EDF $F(<\Mch)$ using modified KS test (KS) and Van der Waerden test (VdW) for the mean sample values. In green shown is the model for which null hypothesis that two distributions have the same mean cannot be rejected at 95\% level (VdW$<1.96$).}
\begin{center}
\begin{tabular}{ |l | l  l  l  l | l  l  l  l  |}
\hline
BH model&\multicolumn{4}{c|}{'CO'}&\multicolumn{4}{c|}{'BH'}\\
\hline
$\alpha_\mathrm{CE}$ & 0.1 & 0.5 & 1 & 4 & 0.1 & 0.5 & 1 & 4 \\ 
 \hline
 \hline
KS & 4.38&	4.22&	3.58&	4.94&	1.70&	1.88&	2.34& 4.30\\ 
VdW & 5.87&	5.03&	5.67&	7.45&\textcolor{green}{1.06}&	2.94 &3.96  &7.09\\
\hline
\end{tabular}
\end{center}
\end{table}

\section{Discussion}
\label{s:discussion}

There are some important points to be discussed. 
\begin{enumerate}[label=(\roman*)]
    \item
    \textbf{Empirical distribution function.}
    The data set we have used to construct the empirical chirp mass distribution of coalescing binary BHs includes (i) 'reliable' sources detected during O1-O2 LIGO/Virgo runs \citep{2019PhRvX...9c1040A}, (ii) less reliable sources found in independent data search in ref. \citep{2020ApJ...891..123N} and (iii) our own estimates of $\Mch$ inferred from the publicly available LIGO/Virgo O3 data based on the assumption on the minimum SNR=8 of the reported O3 detections. To take into account the different reliability of the sources, when constructing EDF $F(<\Mch)$ we have ascribed each source the weight $w_i=\mathrm{SNR}_i/\mathrm{SNR}_{\min}$, where SNR$_{\min}\approx 6$ is the minimum reported SNR in the sample from Table 3 of ref. \citep{2020ApJ...891..123N}. Therefore, thus constructed EDF (shown in blue step line in Figs. \ref{f:KS_EDF}, \ref{f:VdW_EDF} and \ref{f:astro_EDF} can be considered as the left boundary of the real EDF, which can be reliably constructed only after O3 LIGO/Virgo results will be fully processed. 
    The published O3 BH-BH source GW190412 with actual SNR $\rho\approx 19$ \citep{2020arXiv200408342T} and GW190521 with $\rho\approx 15$ \cite{2020arXiv200901190T} show that our estimate of $\Mch$ for O3 sources can be accurate to $\sim 25\%$, considering these BH-BH binaries are among the O3 sources with the highest false alarm rate (see https://gracedb.ligo.org/superevents/public/O3/).
    
    \item
    \textbf{PBH formation model.}
    The PBH formation model we applied \citep{1993PhRvD..47.4244D} predicts a universal log-normal distribution of sizes of high baryon number bubbles created at the inflationary cosmological stage. Outside the horizon, they can be perceived as isocurvature perturbations. After the QCD phase transition at $T\sim 100$~MeV, these perturbations, upon entering the horizon, turn into large density perturbations and form BHs with the log-normal mass spectrum (\ref{e:dN-dM}). As shown in ref. \citep{2020arXiv200411669D}, the mean mass of thus formed PBH distribution should be $\sim 10 M_\odot$, the mass inside the horizon at the QCD phase transition. Therefore, the mean mass of the PBH log-normal distribution derived from our analysis $M_0\simeq 17 M_\odot$ is  close to the expected value.
    \item
    
    \textbf{Binary PBH evolution model.}
    The model \cite{1997ApJ...487L.139N,1998PhRvD..58f3003I} we use catches important aspects of the physics of binary PBH evolution and mergings. However, it was shown to overestimate the PBH merging rate ${\cal R}$ by a suppression factor $C(f_{PBH})<1$ (see, e.g., more sophisticated calculations in Ref. \cite{2017PhRvD..96l3523A,2019JCAP...02..018R}).
    The suppression factor $C$ should appear in our eq. (3.6) and in eq. (3.9) in both numerator and denumerator. It is eq. (3.9) that is important for our calculations of the chirp mass distribution, and the result should not be sensitive to $C$. This guess is supported by a more rigorous analysis in Ref. \cite{2020arXiv200813704H} (see their Fig. 13, green and red curves with and without $C$, respectively). 
    
    \item
    \textbf{Astrophysical models}. 
    In our analysis, we have also compared some particular astrophysical models of binary BH coalescence \citep{2019MNRAS.483.3288P}. Of course, these calculations are based on certain model assumptions, and have been used to construct joint distribution of coalescing binary BH on both chirp mass $\Mch$ and effective spin before the coalescence $\chi_\mathrm{eff}$ 
    Here we have compared only the
    chirp mass distribution $F_\mathrm{PBH}(<\Mch)$ constructed from this model and ignored the effective spin. However, to explain the chirp  mass EDF, we found the deficit of high-mass BHs produced by stellar evolution. This can be due to the BH formation mechanism from massive stars adopted in our models (see, e.g., recent studies \cite{2020arXiv200211278F,
    2020arXiv200400650B,
    2020arXiv200405187V,  2020arXiv200407382B,2020arXiv200409525D} and references therein, for several alternative scenarios). Therefore, our results cannot by no means prohibit astrophysical channels of the binary BH formation but rather strengthen the need to find reliable ways of the formation of massive binary BHs from stellar evolution.
    
    \item
    In the usual scenario of PBH formation at the radiation dominated stage from the collapse of large perturbarions, no significant initial spins are expected
    \citep{2019JCAP...05..018D,2020JCAP...03..017M}. Prior to coalescence, effective PBH spins of binary PBHs can increase due to accretion  \cite{2019JCAP...06..044P,2020JCAP...04..052D}.
    Reliable measurements of large effective spins $\chi_\mathrm{eff}$ in some merging binary BHs (e.g., GW190412 \cite{2020arXiv200408342T}) could be taken as a signature of their non-primordial origin. Apart from the obvious remark that part of coalescing binary BHs could be of different origin, one may also note that a high spin $\sim 0.6$ of one of the components can be acquired during previous merging. The PBH mergings are effective in the possible PBH clusters studied in ref. \cite{2019EPJC...79..246B}. 
   Assuming that PBHs conserve their initial small spins up to the coalescence, 
    ref. \cite{2019JCAP...08..022F} concluded that the effective spin distribution expected from  binary PBH
    mergings does not contradict the available O1-O2 LIGO/Virgo data and can be used to disentangle the possible fraction of primordial and astrophysical binary BHs.
    However, in the case of possible accretion-induced spin-up, massive PBHs could acquire significant spins \cite{2020JCAP...04..052D}, which will be tested in future observations.
    
    \item
    \textbf{Comparison with previous works.}
    In ref. \cite{2019JCAP...02..018R}, parameters of log-normal PBH mass distribution in the form $\psi(M)\sim (1/M)\exp[-\ln^2(M/M_c)/2\sigma^2]$ 
    have been estimated from O1-O2 LIGO/Virgo data. The best-fit model was found to be $M_c=20 M_\odot$, $\sigma=0.6$.
    A similar result with $M_c\approx 17 M_\odot$ was obtained in \cite{2019JCAP...10..059C}. These values correspond to 
    parameters $M_0$ and $\gamma$ of the mass spectrum (\ref{e:dN-dM}) $M_0=M_c e^{-\sigma^2}\approx 14 M_\odot$ $\gamma=1/(2\sigma^2)\approx 1.4$. 
    
    While close to our best-fit values, the difference can be due to
    our using different technique for statistical comparison and extended data sample including estimates of $\Mch$ from public O3 LIGO/Virgo data. We have not addressed the question about the PBH fraction $f_\mathrm{PBH}$ to explain the observed binary BH merging rate. The PBH merging rate, the component mass ratio and chirp mass distribution for best-fit parameters for the model from \cite{2019JCAP...02..018R} were investigated in detail in ref. \cite{2020JCAP...01..031G,2020arXiv200813704H}. 
\end{enumerate}

\section{Conclusion}
\label{s:conclusion}

In this paper, we have checked whether PBH binaries with log-normal mass spectrum (\ref{e:dN-dM}) predicted in ref. \cite{1993PhRvD..47.4244D} can explain the distribution of chirp masses of coalescing binary BHs inferred from LIGO/Virgo observations. To this aim, we have constructed the empirical distribution function $F(<\Mch)$ using data from GWTC-1 catalog \citep{2019PhRvX...9c1040A} supplemented with possible BH-BH mergings found in independent searches in ref. \cite{2020ApJ...891..123N}, as well as estimates of $\Mch$ in O3 sources from public O3 data. Thus constructed in this way EDF was compared with theoretical distributions $F_\mathrm{PBH}(<\Mch)$ calculated using the model of PBH merging rate evolution proposed in ref. \cite{1997ApJ...487L.139N,1998PhRvD..58f3003I} with an account of the actual sensitivity of 
LIGO/Virgo GW detectors. The cumulative distribution $F_\mathrm{PBH}(<\Mch)$ of chirp masses of coalescing binary PBHs that can be registered by a detector with given sensitivity at the fixed signal-to-noise ratio is independent of the unknown fraction $f_\mathrm{PBH}$ of PBHs in cold dark matter.

The modified Kolmogorov-Smirnov test and Van der Waerden non-parametric statistical tests shows that  
the null hypothesis that the observed sample of sources is randomly drawn from the theoretical distribution cannot be rejected at the 95\% level for a range of 
parameters of the log-normal PBH mass distribution (\ref{e:dN-dM})
with $M_0\simeq  13-19 M_\odot$ and $\gamma\simeq 1$ (see Table \ref{t:PBHKS} and \ref{t:PBHVdW}-\ref{t:PBHVdW_var}, respectively). The acceptable model for both tests shown in Fig. \ref{f:KS_EDF} and \ref{f:VdW_EDF} reveals a  good agreement with the observed EDF $F(<\Mch)$. 

Clearly, a more rigorous statistical analysis can be and undoubtedly will be performed once the reliable chirp mass distribution of coalescing binary BHs in O3 LIGO/Virgo run is publicly available. Nevertheless, the result we inferred from our analysis can be suggestive. Indeed, in the considered PBH formation model, the value $M_0\sim 10 M_\odot$ is actually expected from the physical arguments (see \cite{2020arXiv200411669D}), being the mass comprised inside the cosmological horizon at the QCD phase transition at $T\sim 100$~MeV. Thus our finding that the PBH mass distribution with $M_0$ around a dozen solar masses fits the observed chirp mass distribution of coalescing binary BHs can be more than a pure coincidence and  supports the PBH formation mechanism proposed in ref. \cite{1993PhRvD..47.4244D} in 1993. 

\acknowledgments

We thank anonymous referees for constructive criticism and useful notes.
The work of ADD, SP and KP was supported by  the RSF Grant 19-42-02004.
NAM and OSS acknowledge support by the Program of development of M.V. Lomonosov Moscow State University (Leading Scientific School 'Physics of stars, relativistic objects and galaxies'). 



\bibliographystyle{JHEP}
\bibliography{PBH} 
\end{document}